\newcommand{\be}{\begin{eqnarray}}
\newcommand{\ee}{\end{eqnarray}}
\newcommand{\dia}{\!\!\!\!\!\not\,\,\,}
\documentclass[twocolumn,aps,showpacs]{revtex4}
\usepackage{graphics}
\begin{document}
\title{Axial charge segregation and $CP$ violation during the
       electroweak phase transition with hypermagnetic fields}  
\author{Alejandro Ayala$^\dagger$, Jaime Besprosvany$^\ddagger$}  
\affiliation{$\dagger$Instituto de Ciencias Nucleares, Universidad Nacional 
         Aut\'onoma de M\'exico, Apartado Postal 70-543, 
         M\'exico Distrito Federal 04510, M\'exico.\\
         $^\ddagger$Instituto de F\'{\i}sica,
         Universidad Nacional Aut\'onoma de M\'exico
         Apartado Postal 20-364, 
         M\'exico Distrito Federal 01000, M\'exico.}
\begin{abstract}
We study the scattering of fermions off the true vacuum bubbles
nucleated during a first order electroweak phase transition in the
presence of a background hypermagnetic field. By considering the limit
of an infinitely thin bubble wall, we derive and analytically solve 
the Dirac equation in three spatial dimensions for left and right chirality
modes and compute reflection and transmission coefficients for the
case when the fermions move from the symmetric toward the broken symmetry
phase. Given the chiral nature of the fermion coupling with the
background field in the symmetric phase, an axial asymmetry is
generated during the scattering processes. This mechanism is thus
shown to provide an additional source of $CP$ violation within the
standard model. We discuss possible implications for baryon number
generation.
\end{abstract}


\maketitle

\section{Introduction}\label{I}

It is well known that the standard model (SM) of electroweak
interactions meets all the requirements --known as Sakharov
conditions~\cite{Sakharov}-- to generate a baryon asymmetry
during the electroweak phase transition (EWPT), provided that this last
be of first order. As appealing as this scenario might be, it is also
well known that neither the amount of $CP$ violation within the minimal
SM nor the strength of the EWPT are enough to generate a sizable
baryon number~\cite{{Gavela},{Kajantie}}. 

Nonetheless, it has been recently shown that in the presence of
large scale primordial magnetic fields it is possible to generate a
stronger first order EWPT~\cite{{Giovannini},{Elmfors},{Giovannini2}}. 
The situation is similar to what happens to a type I superconductor
where an external magnetic field modifies the nature of the
superconducting phase transition, changing it from second to first
order due to the Meissner effect.  

The presence of magnetic fields during the evolution of the early
universe has neither been proved nor has it been ruled
out. Temperature anisotropies measured by COBE place an upper bound
$B_0\sim 10^{-9}\ $G for homogeneous fields ($B_0$ refers to the
intensity that the field would have today under the assumption of
adiabatic decay due to the Hubble expansion)~\cite{Barrow}. In the
case of inhomogeneous fields their effect must be searched for in the
Doppler peaks~\cite{Adams} and in the polarization of the cosmic
microwave background radiation~\cite{Kosov}. The future satellite
missions MAP and PLANCK may reach the required sensitivity for the
detection of these last signals. 

Independently of their origin, there are several interesting cosmological
consequences of the possible existence of primordial magnetic fields
such as the generation of galactic seed fields and their influence
on big bang nucleosynthesis~\cite{reviews}. 

Recall that for temperatures above the EWPT, the SU(2)$\times$U(1)$_Y$
symmetry is restored and the propagating, non-screened vector modes that
represent a magnetic field correspond to the U(1)$_Y$
group instead of to the U(1)$_{em}$ group, and are therefore properly
called {\it hypermagnetic} fields. 

In previous works~\cite{Ayala2}, it has been shown that another
interesting cosmological consequence of the presence of such fields
during the EWPT is the production of an axial charge segregation in
the scattering of fermions off the true vacuum bubbles within the
SM. This charge segregation is a consequence of the chiral  
nature of the fermion coupling to hypermagnetic fields in the
symmetric phase. The analysis relied on the solution of an effectively
one-dimensional Dirac equation for fermions moving perpendicular to
the bubble wall. 

In this paper, we lift this restriction and solve the Dirac equation
in three spatial dimensions for the two chirality fermion modes in 
the presence of an external hypermagnetic field. Working in the limit
of an infinitely thin bubble wall, we find analytical solutions 
--valid for fermion wave functions described by small quantum numbers-- from
where we compute reflection and transmission coefficients and show
explicitly that these differ for left and right-handed chirality
modes incident from the symmetric phase. Since these coefficients are
related to the ones obtained by studying fermion incidence from the
broken symmetry phase by $CPT$ and unitarity, we show how an axial charge
segregation is built during the process. Such axial charge segregation
provides a bias for baryon over antibaryon production and thus a seed
for the non-local baryogenesis
scenario~\cite{{Dine},{Cohen},{Nelson}}. We also discuss how this
mechanism is equivalent to an extra source of $CP$ violation.   

The outline of this work is as follows: In Sec.~\ref{II}, we write
the Dirac equation for the left and right-handed chirality
modes propagating in a background hypermagnetic field during the
EWPT where the spatial profile of the Higgs field is taken in the
limit of an infinitely thin wall. In
Sec.~\ref{III}, we find the three dimensional solutions to this
equation and discuss their properties. In
Sec.~\ref{IV}, we use the solutions 
to compute reflection and transmission coefficients for axial fermion modes
moving from the symmetric phase toward the broken symmetry
phase. Finally in Sec.~\ref{V}, we conclude by looking out at the 
possible implications of such axially asymmetric fermion reflection
and transmission. 

\section{Dirac equation for chiral fermions in a background
hypermagnetic field}\label{II}

During a first order EWPT, the properties of the nucleated bubbles
depend on the effective, finite temperature Higgs potential. In the
{\it thin wall} approximation and under the assumption that the phase
transition happens when the energy densities of both phases are
degenerate, it is possible to find a one-dimensional analytical
solution for the Higgs field spatial profile $\phi (z)$ called the {\it
kink}. This is given by 
\be
   \phi (z) \sim 1 + \tanh (z/\lambda)\, ,
   \label{kink}
\ee
where $z$ is the coordinate along the direction of the phase change and
$\lambda$ is the width of the wall. 

When $\lambda$ is small compared to the mean free path, scattering
within the wall is not affected by diffusion and the problem of
fermion reflection and transmission through the wall can be cast in
terms of solving the Dirac equation with a position dependent fermion mass,
proportional to the Higgs field~\cite{Ayala}, where the
proportionality constant is given by the particle's Yukawa coupling. 
In physical terms, the wall can be considered thin when the spatial region
where the fermion mass changes is small compared to other
relevant length scales such as the particle mean free path.

In order to obtain analytical solutions, let us further simplify the
problem considering the limit of an infinitely thin bubble wall,
namely $\lambda\rightarrow 0$. In this case, the kink solution becomes
a step function, $\Theta (z)$, and consequently, the expression for
the particle's mass becomes  
\be
   m(z)=m_0\Theta (z)\, .
   \label{step}
\ee   
For $z\leq 0$ particles move in the region outside the bubble, that is,
in the symmetric phase where particles are massless. Conversely, for
$z\geq 0$, particles move inside the bubble, which represents the
broken phase where particles have acquired a finite mass $m_0$.

In the presence of an external magnetic field, we need to consider
that fermion modes couple differently to the field in the broken and
the symmetric phases. 

For $z\leq 0$, the coupling is chiral. Let 
\be
   \Psi_R&=&\frac{1}{2}\left(1 + \gamma_5\right)\Psi\nonumber\\
   \Psi_L&=&\frac{1}{2}\left(1 - \gamma_5\right)\Psi
   \label{chiralmodes}
\ee
represent, as usual, the right and left-handed chirality modes for the
spinor $\Psi$, respectively. Then, the equations of motion for these
modes, as derived from the electroweak interaction Lagrangian, are
\be
   (i\partial\dia\ -\ \frac{y_L}{2}g'A\dia\ )
   \Psi_L - m(z)\Psi_R &=& 0\nonumber\\ \nonumber\\
   (i\partial\dia\ -\ \frac{y_R}{2}g'A\dia\ )
   \Psi_R - m(z)\Psi_L &=& 0\, ,
   \label{diracsymm}
\ee 
where $y_{R,L}$ are the right and left-handed hypercharges
corresponding to the given fermion, respectively, $g'$ the
$U(1)_Y$ coupling constant and we take $A^\mu=(0,{\mathbf A})$
representing a, not yet specified, four-vector potential having
non-zero components only for its transverse spatial part, in the rest
frame of the wall. This last requirement means that the corresponding
magnetic field will point along the direction perpendicular to the
wall. 

The gauge field profile should correspond to our choice of Higgs
field profile and thus the change in the magnetic field strength
should take place in the same small spatial region over which the change in
the Higgs field takes place. An exact treatment that considered a
continuous change in the Higgs field would require that the gauge
field be also continuous across the interface. Nonetheless, since we
are pursuing insight from analytical results, we consider the
simplified situation in which the gauge field is constant on either
side of the wall and discontinuous at the planar phase boundary. A
discussion on the consequences of considering a continuous gauge field
profile can be found in the second of Refs.~\cite{Ayala2}.

The set of Eqs.~(\ref{diracsymm}) can be written as a single equation
for the spinor $\Psi = \Psi_R + \Psi_L$ by adding them up
\begin{widetext}
\be
   \left\{ i\partial\dia\ - A\dia
   \left[\frac{y_R}{4}g'\left(1 + \gamma_5\right)
   +\frac{y_L}{4}g'\left(1 - \gamma_5\right)\right]
   - m(z)\right\}\Psi = 0\, .
   \label{diracsingle}
\ee
\end{widetext}
Hereafter, we explicitly work in the chiral representation of the
gamma matrices where
\be
   \gamma^0=\!\left(\begin{array}{rr}
   0 & -I \\
   -I & 0 \end{array}\right)\
   \mbox{\boldmath $\gamma$}=\!\left(\begin{array}{rr}
   0 &  \mbox{\boldmath $\sigma$} \\
   \mbox{\boldmath $-\sigma$} & 0 \end{array}\right)\
   \gamma_5=\!\left(\begin{array}{rr}
   I & 0 \\
   0 & -I \end{array}\right)\, .
   \label{gammaschiral}
\ee
Within this representation, we can write Eq.~(\ref{diracsingle}) as
\be
   \Big\{i\partial\dia\ -\ {\mathcal G}A_\mu\gamma^\mu
   -m(z)\Big\}\Psi=0\, ,
   \label{diracsimple}
\ee
where we have introduced the matrix
\be
   {\mathcal G}=\left(\begin{array}{cc}
   \frac{y_L}{2}g'I & 0 \\
   0 & \frac{y_R}{2}g'I \end{array}\right)\, .
   \label{matA}
\ee
We now look at the corresponding equation in the broken symmetry 
phase. For $z\geq 0$ the coupling of the fermion with the external
field is through the electric charge $e$ and thus, the equation of motion
is simply the Dirac equation describing an electrically
charged fermion in a background magnetic field, namely,
\be
   \Big\{i\partial\dia\ -\ eA_\mu\gamma^\mu
   -m(z)\Big\}\Psi=0\, .
   \label{diracsimplezg0}
\ee
In the following section, we explicitly construct the solutions to
Eqs.~(\ref{diracsimple}) and ~(\ref{diracsimplezg0}) with a constant
magnetic field, requiring that these match at the interface $z=0$.

\section{Solving the Dirac Equation}\label{III}

For definiteness, let us consider a constant magnetic field
${\mathbf B}=B\hat{z}$ pointing along the $\hat{z}$ direction. We
first look for the stationary solutions of Eq.~(\ref{diracsimple}) 
corresponding to fermions moving in the symmetric phase. Writing
\be
   \Psi ({\bf r},t)=e^{-iEt}\Psi ({\bf r})\, ,
   \label{stationary}
\ee 
where
\be
   \Psi ({\bf r}) = \left(
                    \begin{array}{c}
                       \psi_1({\bf r})\\
                       \psi_2({\bf r})\\
                       \psi_3({\bf r})\\
                       \psi_4({\bf r})
                    \end{array}\right)\, ,
   \label{compon}
\ee
and working in the chiral representation of the gamma matrices,
Eq.~(\ref{diracsimple}) becomes explicitly 
\be
   E\psi_1({\bf r}) - (P_x^R - iP_y^R)\psi_2({\bf r}) 
   - P_z\psi_1({\bf r}) &=&0\nonumber\\
   E\psi_2({\bf r}) - (P_x^R + iP_y^R)\psi_1({\bf r}) 
   + P_z\psi_2({\bf r}) &=&0\nonumber\\
   E\psi_3({\bf r}) + (P_x^L - iP_y^L)\psi_4({\bf r}) 
   + P_z\psi_3({\bf r}) &=&0\nonumber\\
   E\psi_4({\bf r}) + (P_x^L + iP_y^L)\psi_3({\bf r}) 
   - P_z\psi_4({\bf r}) &=&0\, ,
   \label{firstsys}
\ee
where we introduced the definitions 
\be
   {\mathbf P}^R &\equiv& -i\nabla + \frac{y^Rg'}{2}{\mathbf
   A}\nonumber\\
   {\mathbf P}^L &\equiv& -i\nabla + \frac{y^Lg'}{2}{\mathbf
   A}\, .
   \label{Mom}
\ee
Since $A_z=0$, 
\be
   P_z^R=P_z^L\equiv P_z\, .
   \label{Momz}
\ee
In this representation $\psi_{1,2}$ and $\psi_{3,4}$ correspond to
components with right- and left-handed chirality, respectively. Notice
that, since in the symmetric phase fermions are massless, the
differential Eqs.~(\ref{firstsys}) do not mix components with
different chirality.

Let us now work in cylindrical coordinates and look for solutions of
the form
\be
   \Psi(r,\phi, z)\sim \left(
                    \begin{array}{c}
                       f_1(r)e^{-i\phi/2}\\
                       f_2(r)e^{+i\phi/2}\\
                       f_3(r)e^{-i\phi/2}\\
                       f_4(r)e^{+i\phi/2}\\
                    \end{array}\right)
   e^{i(l-\frac{1}{2})\phi}e^{ikz}\, .
   \label{eigenzphi}
\ee
Using the explicit components of the vector ${\bf A}({\bf r})$ that
produce a constant magnetic field along the $\hat{z}$ axis, it is easy
to show that~\cite{Sokolov}
\be
   P_x^{R,L}\pm iP_y^{R,L}=-ie^{\pm i\phi}
   \left(\frac{\partial}{\partial r}\pm \frac{i}{r}
   \frac{\partial}{\partial\phi}\mp\gamma^{R,L}r\right)\, ,
   \label{P'scil}
\ee
which, together with the definitions
\be
   \gamma^{R,L}&\equiv& \frac{y^{R,L}g'}{4}B\nonumber\\
   \rho^{R,L}&\equiv&\gamma^{R,L}r^2\nonumber\\
   {\cal R}^{R,L}_1&\equiv& (\gamma^{R,L}\rho^{R,L})^{1/2}
   \left( 2\frac{d}{d\rho^{R,L}} - 1 - 
   \frac{(l-1)}{\rho^{R,L}}\right)\nonumber\\
   {\cal R}^{R,L}_2&\equiv& (\gamma^{R,L}\rho^{R,L})^{1/2}
   \left( 2\frac{d}{d\rho^{R,L}} + 1 + 
   \frac{l}{\rho^{R,L}}\right)\, ,
   \label{defs}
\ee
make the set of Eqs.(\ref{firstsys}) to be explicitly written as
\be
   (E-k)f_1(\rho^R)+i{\cal R}^R_2f_2(\rho^R)&=&0\nonumber\\
   (E+k)f_2(\rho^R)+i{\cal R}^R_1f_1(\rho^R)&=&0\nonumber\\
   (E+k)f_3(\rho^L)+i{\cal R}^L_2f_4(\rho^L)&=&0\nonumber\\
   (E-k)f_4(\rho^L)+i{\cal R}^L_1f_3(\rho^L)&=&0\, .
   \label{secondsys}
\ee
By acting on Eqs.~(\ref{secondsys}) with $i{\cal R}^R_1$, $i{\cal
R}^R_2$, $i{\cal R}^L_1$ and $i{\cal R}^L_2$, respectively, we obtain
the set of second order differential equations
\begin{widetext}
\be
   \left\{\rho^R\frac{d^2}{d(\rho^R)^2} + \frac{d}{d\rho^R} +
   \lambda^R - \frac{l}{2} - \frac{\rho^R}{4} - 
   \frac{(l-1)^2}{4\rho^R}\right\}f_1(\rho^R)&=&0\nonumber\\
   \left\{\rho^R\frac{d^2}{d(\rho^R)^2} + \frac{d}{d\rho^R} +
   \lambda^R - \frac{(l-1)}{2} - \frac{\rho^R}{4} - 
   \frac{l^2}{4\rho^R}\right\}f_2(\rho^R)&=&0\nonumber\\
   \left\{\rho^L\frac{d^2}{d(\rho^L)^2} + \frac{d}{d\rho^L} +
   \lambda^L - \frac{l}{2} - \frac{\rho^L}{4} - 
   \frac{(l-1)^2}{4\rho^L}\right\}f_3(\rho^L)&=&0\nonumber\\
   \left\{\rho^L\frac{d^2}{d(\rho^L)^2} + \frac{d}{d\rho^L} +
   \lambda^L - \frac{(l-1)}{2} - \frac{\rho^L}{4} - 
   \frac{l^2}{4\rho^L}\right\}f_4(\rho^L)&=&0\, ,
   \label{secordsys}
\ee
\end{widetext}
where
\be
   \lambda^{R,L}\equiv\frac{E^2-(k^{R,L})^2}{4\gamma^{R,L}}\, .
   \label{lambd}
\ee
The eigenvalues $\lambda^{R,L}$ corresponding to the requirement that
the solutions vanish as $\rho^{R,L}\rightarrow\infty$ are such that
\be
   \lambda^{R,L}=n\, .
   \label{n}
\ee
$n=l+s$ is called the principal quantum number which must be a
non-negative integer $n=0,1,2,3\ldots$ From Eq.~(\ref{lambd}), the
spectrum is given by
\be
   E=\sqrt{(k^{R,L})^2+4n\gamma^{R,L}}\, .
   \label{spectrumsym}
\ee 
The solutions to Eqs.~(\ref{secordsys}) are
\be
   f_1(\rho^R)&=& C_1I_{n-1,s}(\rho^R)\nonumber\\
   f_2(\rho^R)&=&iC_2I_{n,s}(\rho^R)\nonumber\\
   f_3(\rho^L)&=& C_3I_{n-1,s}(\rho^L)\nonumber\\
   f_4(\rho^L)&=&iC_4I_{n,s}(\rho^L)\, ,
   \label{sol}
\ee
where $C_{1,\ldots,4}$ are constants and $I_{n,s}(\rho)$ are the
Laguerre functions given in terms of the Laguerre polynomials
$Q_s^{n-s}$ by 
\be
   I_{n,s}(\rho)=\frac{1}{(n!s!)^{1/2}}e^{-\rho/2}\rho^{(n-s)/2}
   Q_s^{n-s}(\rho)\, ,
   \label{Laguerre}
\ee
satisfying the normalization condition
\be
   \int_0^\infty d\rho I_{n,s}^2(\rho)=1\, .
   \label{norm}
\ee
Using the recurrence relations for the Laguerre polynomials together
with the set of Eqs.~(\ref{firstsys}), one readily obtains that the
constants $C_{1,2}$ and $C_{3,4}$ are not independent and thus that
the solution $\Psi({\bf r})$ for fermions moving in the symmetric
phase is given by
\begin{widetext}
\be
   \Psi(r,\phi, z)= \left(
                    \begin{array}{c}
                       C_2\left\{
                       \begin{array}{c}
                          \sqrt{\frac{E+k^R}{E-k^R}}
                          I_{n-1,s}(\rho^R)e^{i(l-1)\phi}\\
                          iI_{n,s}(\rho^R)e^{il\phi}
                       \end{array}\right\}e^{ik^Rz}\\
                       C_3\left\{
                       \begin{array}{c}
                          I_{n-1,s}(\rho^L)e^{i(l-1)\phi}\\
                          i\sqrt{\frac{E+k^L}{E-k^L}}
                          I_{n,s}(\rho^L)e^{il\phi}
                       \end{array}\right\}e^{ik^Lz}
                    \end{array}\right)\, .
   \label{finalpsisym}
\ee
\end{widetext}
We now turn to finding the solutions to Eq.~(\ref{diracsimplezg0}),
namely, for fermions moving in the broken symmetry phase. By a
procedure similar to the above, it is easy to see that the solutions
are given by~\cite{Sokolov}
\be
   \Psi(r,\phi, z)= \left(
                    \begin{array}{l}
                       D_1I_{n'-1,s'}(\rho)e^{i(l'-1)\phi}\\
                       iD_2I_{n',s'}(\rho)e^{il'\phi}\\
                       D_3I_{n'-1,s'}(\rho)e^{i(l'-1)\phi}\\
                       iD_4I_{n',s'}(\rho)e^{il'\phi}
                       \end{array}
                    \right)e^{ikz}\, ,
   \label{finalpsibrok}
\ee
where
\be
   \rho&\equiv&\gamma r^2\nonumber\\
   \gamma&\equiv& \frac{g'}{2}B\, .
   \label{defsbrok}
\ee
Notice that in the second of Eqs.~(\ref{defsbrok}), we have used that
since in the broken symmetry phase there should not be a propagating
component corresponding to the $Z^0$ field~\cite{Giovannini,Elmfors},
the parameter representing the magnetic field strength $B'$ is related
to $B$ and Weinberg's angle $\theta_W$ by
\be
   B'=\frac{B}{\cos\theta_W}\, ,
   \label{relat}
\ee
which in turn implies that the coupling of the fermion with the
magnetic field in the broken symmetry phase $eB'$ becomes 
\be
   eB'=g'B\, .
   \label{implies}
\ee
The spectrum is given by $E=\sqrt{k^2+m_0^2+4n'\gamma}$, with $n'=l'+s'$ being
a non-negative integer $n'=0,1,2\ldots$. The constants $D_{1,\ldots,4}$
are not independent from each other and are given in terms of the two
constants $G_1$ and $G_2$ by
\begin{widetext}
\be
   \left(\begin{array}{c}
            D_1\\
            D_2\\
            D_3\\
            D_4
         \end{array}\right) = 
    G_1\left(\begin{array}{c}
                -\frac{\sqrt{E^2-k^2-m_0^2}}{m_0}\\
                -\frac{(E-k)}{m_0}\\
                0\\
                1
             \end{array}\right) +
    G_2\left(\begin{array}{c}
                -\frac{(E+k)}{m_0}\\
                -\frac{\sqrt{E^2-k^2-m_0^2}}{m_0}\\
                1\\
                0
             \end{array}\right)\, .
   \label{D's}
\ee
\end{widetext}
In order to describe the motion of fermions in the whole domain, we
should now proceed to join the solutions in Eqs.~(\ref{finalpsisym})
and~(\ref{finalpsibrok}) across the interface at $z=0$. To this end, we
work in the approximation where the incident fermions from the
symmetric phase make energy and angular momentum conserving
transitions to states with the same principal
quantum number in the broken symmetry phase, namely $n=n'$ and
$l=l'$, which requires $s=s'$. This approximation corresponds to the
classical picture in which the effect of a sudden change in the
coupling with the external field produces only a change in the radius of
the orbit around the field lines. This is a reasonable approximation as
long as the number of quanta in the transverse modes for a given
energy is not too large. This can be quantified by considering the
overlap of the radial functions $I_{n,s}(\rho^{R,L})$ and
$I_{n',s'}(\rho)$ for $l=l'$ as a function of $s'-s$, as shown in
Fig.~1. We find that our approximation is valid for low values of $s$
and $l$. Otherwise, interference with modes with additional nodes
associated with large values of these quantum numbers is not
negligible, which means these modes need to be considered in the
complete solution. 

For fermions incident from the symmetric phase, the complete wave
function contains incoming as well as outgoing components, whereas in
the broken symmetry phase the wave function corresponds to only
outgoing waves. To describe the scattering of right- and left-handed
chirality modes, let us consider them separately.

For right-handed incident modes, continuity of the solution across the
interface yields the system of conditions
\begin{widetext}
\be
   \sqrt{\frac{E+k^R}{E-k^R}} + \left(\sqrt{\frac{E-k^R}{E+k^R}}\ \right)
   C_2^R &=& - \left(\frac{\sqrt{E^2-k^2-m_0^2}}{m_0}\ \right) G_1^R
   - \left(\frac{E+k}{m_0}\right) G_2^R\nonumber\\
   1+C_2^R&=&- \left(\frac{E-k}{m_0}\right) G_1^R
   -\left(\frac{\sqrt{E^2-k^2-m_0^2}}{m_0}\ \right) G_2^R\nonumber\\
   C_3^R&=&G_2^R\nonumber\\
   \left(\sqrt{\frac{E-k^L}{E+k^L}}\ \right) C_3^R &=& G_1^R\, .
   \label{sysright}
\ee
\end{widetext}
On the other hand, for left-handed incident modes, continuity of the
solution across the interface yields the system of conditions
\begin{widetext}
\be
   \left(\sqrt{\frac{E-k^R}{E+k^R}}\ \right)
   C_2^L &=& - \left(\frac{\sqrt{E^2-k^2-m_0^2}}{m_0}\ \right) G_1^L
   - \left(\frac{E+k}{m_0}\right) G_2^L\nonumber\\
   C_2^L&=&- \left(\frac{E-k}{m_0}\right) G_1^L
   -\left(\frac{\sqrt{E^2-k^2-m_0^2}}{m_0}\ \right) G_2^L\nonumber\\
   1+C_3^L&=&G_2^L\nonumber\\
   \left(\sqrt{\frac{E+k^L}{E-k^L}}\ \right) +
   \left(\sqrt{\frac{E-k^L}{E+k^L}}\ \right) C_3^L &=& G_1^L\, .
   \label{sysleft}
\ee
\end{widetext}
Eqs.~(\ref{sysright}) and~(\ref{sysleft}) completely determine the
constants $C_{2,3}^{R}$, $G_{1,2}^R$ and $C_{2,3}^{L}$, $G_{1,2}^L$,
respectively. 

\section{Transmission and reflection coefficients}\label{IV}

\begin{figure}[b] 
{\centering{
\resizebox*{0.4\textwidth}{!}
{\includegraphics{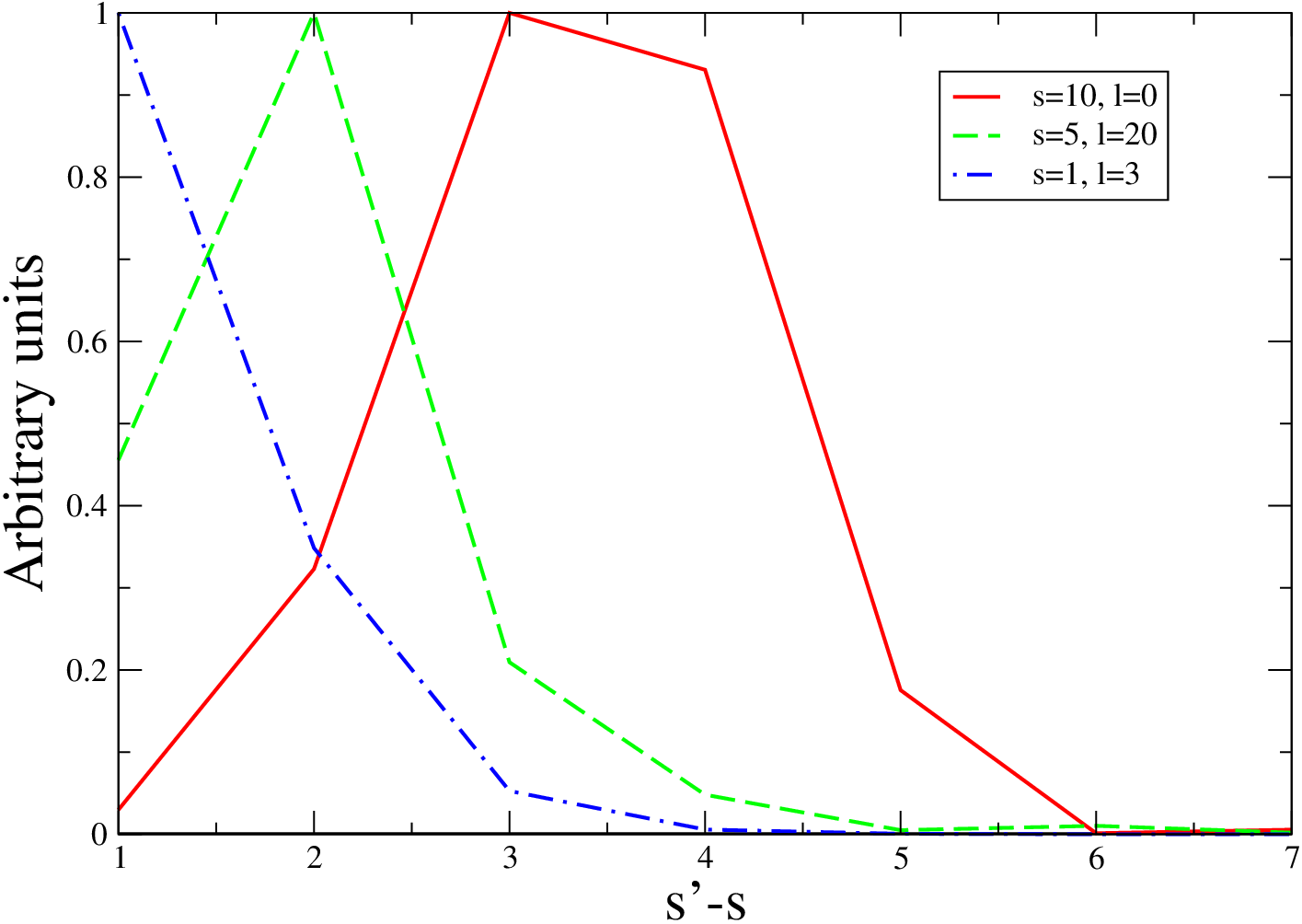}}}\par} 
\caption{Overlap integral for the functions $I_{n,s}(\rho^{R})$ and
$I_{n',s'}(\rho)$ normalized to the maximum value in each case for
$l=l'$ as a function of $s'-s$ for three 
different values of the pairs $s$ and $l$. Notice that when $s$ and
$l$ are large, the overlap is maximum for $s'\neq s$.} 
\end{figure}

The fact that the solutions to the systems of Eqs.~(\ref{sysright})
and~(\ref{sysleft}) are not the same, means that there is the
possibility of building an axial asymmetry during the scattering of
fermions off the wall. To quantify the asymmetry, we need to compute
the corresponding reflection and transmission coefficients. These are
built from the reflected, transmitted and incident currents of each
type. Recall that for a given spinor wave function $\Psi$, the current
normal to the wall is given by 
\be
   J=\int_0^\infty d\rho\int_0^{2\pi}d\phi
   \Psi^\dagger(\rho,\phi,z)\gamma^0\gamma^3\Psi(\rho,\phi,z)\, .
   \label{current}
\ee
For right-handed incoming waves, the incident current $J^R_{\mbox{\small
inc}}$ is thus given by  
\be
   J^R_{\mbox{\small inc}}=\frac{2k^R}{E-k^R}
   \label{jRinc}
\ee
whereas the reflected and transmitted currents $J^R_{\mbox{\small
ref}}$, $J^R_{\mbox{\small tra}}$ are given respectively by
\be
   J^R_{\mbox{\small ref}}&=&\left(\frac{2k^R}{E+k^R}\right)(C_2^R)^2
   + \left(\frac{2k^L}{E+k^L}\right)(C_3^R)^2
   \nonumber\\
   J^R_{\mbox{\small tra}}&=&\left[
   \left(\frac{\sqrt{E^2-k^2-m_0^2}}{m_0}\ \right) G_1^R +
   \left(\frac{E+k}{m_0}\right) G_2^R\right]^2\nonumber\\
   &-&
   \left[
   \left(\frac{\sqrt{E^2-k^2-m_0^2}}{m_0}\ \right) G_2^R +
   \left(\frac{E-k}{m_0}\right) G_1^R\right]^2\nonumber\\
   &-&
   (G_2^R)^2 + (G_1^R)^2\, .
   \label{jrreftrans}
\ee
On the other hand, for left-handed incoming waves, the incident
current $J^L_{\mbox{\small inc}}$ is given by
\be
   J^L_{\mbox{\small inc}}=\frac{2k^L}{E-k^L}
   \label{jLinc}
\ee
and the reflected and transmitted currents $J^L_{\mbox{\small
ref}}$, $J^L_{\mbox{\small tra}}$ are given respectively by
\be
   J^L_{\mbox{\small ref}}&=&\left(\frac{2k^R}{E+k^R}\right)(C_2^L)^2
   + \left(\frac{2k^L}{E+k^L}\right)(C_3^L)^2
   \nonumber\\
   J^L_{\mbox{\small tra}}&=&\left[
   \left(\frac{\sqrt{E^2-k^2-m_0^2}}{m_0}\ \right) G_1^L +
   \left(\frac{E+k}{m_0}\right) G_2^L\right]^2\nonumber\\
   &-&
   \left[
   \left(\frac{\sqrt{E^2-k^2-m_0^2}}{m_0}\ \right) G_2^L +
   \left(\frac{E-k}{m_0}\right) G_1^L\right]^2\nonumber\\
   &-&
   (G_2^L)^2 + (G_1^L)^2\, .
   \label{jlreftrans}
\ee
Recall that in the symmetric phase, where fermions are massless, the
chirality and helicity operators commute and thus chirality modes are
also eigenfunctions of helicity. From Eq.~(\ref{finalpsisym}) and the
representation of the gamma matrices, Eq.~(\ref{gammaschiral}), we see
that for right-(left) handed chirality modes, the large components
correspond to right-(left) handed helicity modes. Since scattering of
the wall does not change the spin direction of the impinging particle,
right-(left) handed helicity modes reflect as left-(right) handed
and transmit as right-(left) handed modes. To
emphasize this point, for the reflection and transmission
coefficients, let us denote by lower case letters $r$ and $l$ the
right- and left-handed helicity modes. These coefficients are given as
the ratios of the corresponding reflected and transmitted currents, to
the incident one, explicitly 
\be
   R_{r\rightarrow l}&=&
   J_{\mbox{\small ref}}^R/
   J_{\mbox{\small inc}}^R\nonumber\\
   T_{r\rightarrow r}&=&
   J_{\mbox{\small tra}}^R/
   J_{\mbox{\small inc}}^R\, .
   \label{RTr}
\ee
The corresponding coefficients for the axially conjugate process are
\be
   R_{l\rightarrow r}&=&
   J_{\mbox{\small ref}}^L/
   J_{\mbox{\small inc}}^L\nonumber\\
   T_{l\rightarrow l}&=&
   J_{\mbox{\small tra}}^L/
   J_{\mbox{\small inc}}^L\, .
   \label{RTl}
\ee
Using Eqs.~(\ref{jRinc}) to~(\ref{jrreftrans}), together with the
solutions to Eqs.~(\ref{sysright}) and~(\ref{sysleft}), it is
straightforward to show that
\be
   R_{r\rightarrow l} + T_{r\rightarrow r} &=& 1\nonumber\\
   R_{l\rightarrow r} + T_{l\rightarrow l} &=& 1
   \label{unitarity}
\ee
Figure~2 shows the coefficients $R_{l\rightarrow r}$ and
$R_{r\rightarrow l}$ as a function of the magnetic field $B$ scaled by
$m_0^2$ for $E/m_0=3$, $n=1$ and hypercharge values for top quarks
$y_R=4/3$, $y_L=1/3$, with $g'=0.344$, as appropriate for
the EWPT epoch. Notice that when $b\rightarrow 0$, these coefficients
approach each other and that the difference grows with increasing
field strength. 

\begin{figure}[t] 
\vspace{-0.8cm}
{\centering
\resizebox*{0.4\textwidth}{!}
{\includegraphics{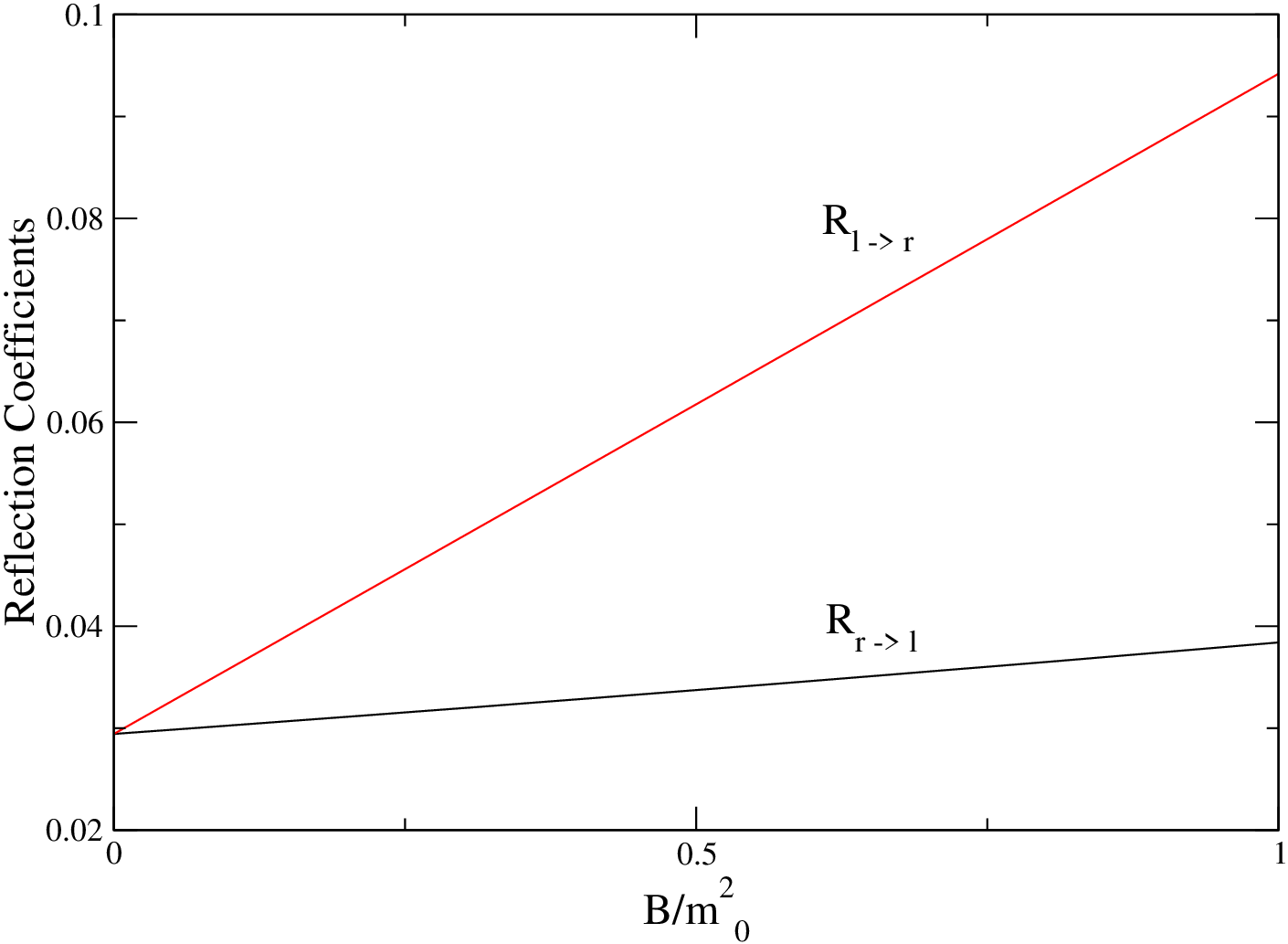}}\par} 
\caption{Coefficients $R_{l\rightarrow r}$ and
$R_{r\rightarrow l}$ as a function of the magnetic
field $B$ scaled by $m_0^2$ for $E/m_0=3$, $n=1$, and $y_R=4/3$,
$y_L=1/3$. The value for the $U(1)_Y$ coupling constant is taken as
$g'=0.344$, corresponding to the EWPT epoch.} 
\end{figure}

Figure~3 shows the reflection and transmission coefficients as a
function of energy scaled by the fermion mass $m_0$. Figure~3a shows the
coefficients $R_{r\rightarrow l}$ and $T_{r\rightarrow r}$ and Fig.~3b
the coefficients $R_{l\rightarrow r}$ and $T_{l\rightarrow l}$ for
$B/m_0^2=0.5$ and $n=1$, $y_R=4/3$, $y_L=1/3$, $g'=0.344$. Since the
solutions in Eqs.~(\ref{finalpsibrok}) are computed assuming that the
transmitted waves are not exponentially damped, their energies have to be
taken such that $E\geq\sqrt{m_0^2+4n\gamma}$. For these energy values,
there is thus no need to consider the contribution from negative-energy
solutions~\cite{Bjorken}. 

\section{Discussion and Conclusions}\label{V}

In this paper we have derived and solved the Dirac equation describing 
fermions scattering off a first order EWPT bubble wall in three
spatial dimensions, in the approximation of an infinitely thin wall,
in the presence of a magnetic field directed along the fermion
direction of motion. In the symmetric phase, the fermions couple
chirally to the magnetic field, 
which receives the name of {\it hypermagnetic}, given that it
belongs to the $U(1)_Y$ group. We have shown that the chiral nature of
this coupling implies that it is possible to build an axial
asymmetry during the scattering of fermions off the
wall. We have computed reflection and transmission coefficients
showing explicitly that they differ for left and right-handed incident
particles from the symmetric phase. The results of this calculation
are in agreement with those previously found in Refs.~\cite{Ayala2},
where the motion of the fermions was effectively treated as
one-dimensional.  

\begin{figure}[t] 
\vspace{-0.8cm}
{\centering
\resizebox*{0.4\textwidth}{!}
{\includegraphics{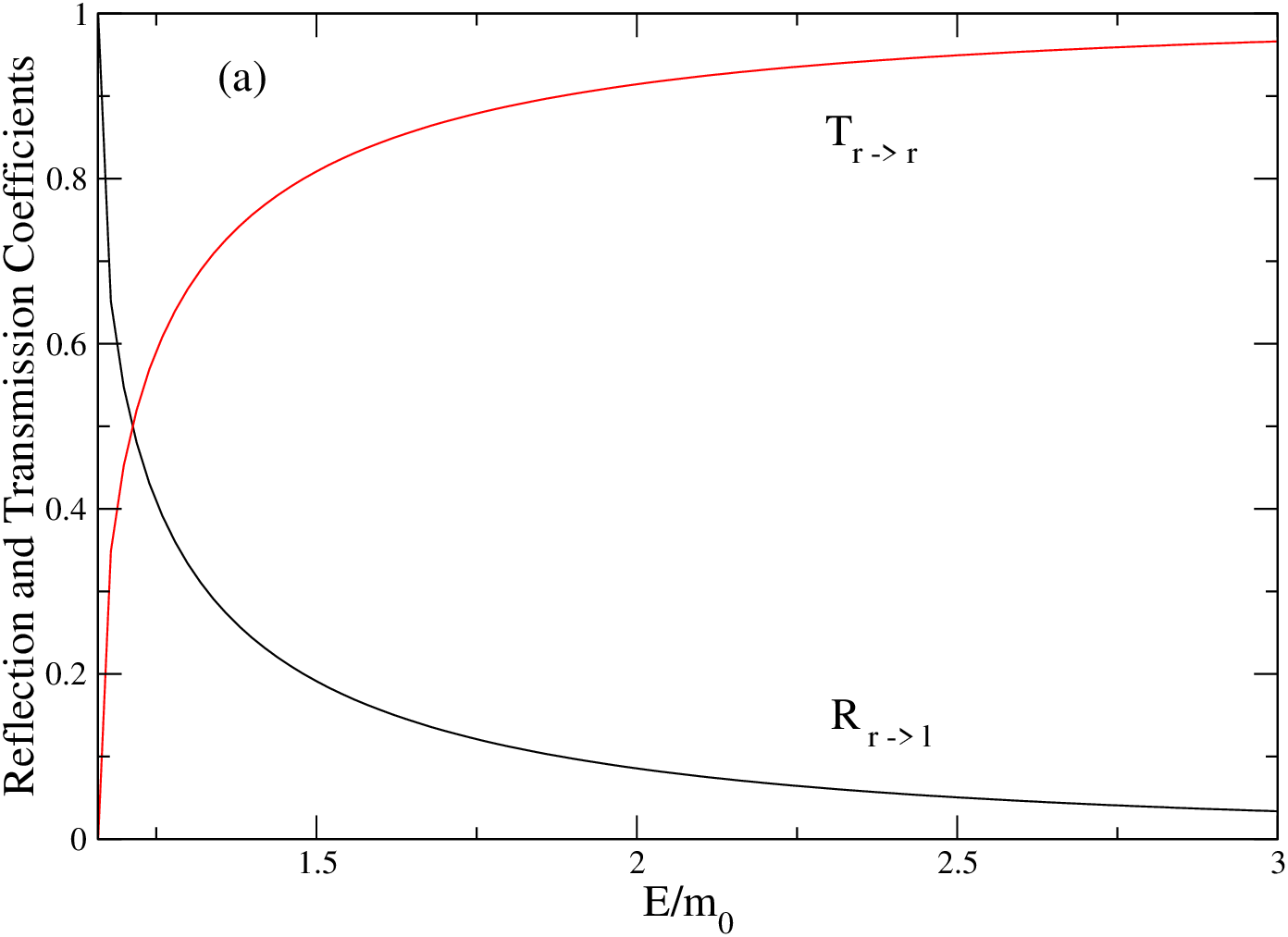}}\par} 
\vspace{1cm}
{\centering
\resizebox*{0.4\textwidth}{!}
{\includegraphics{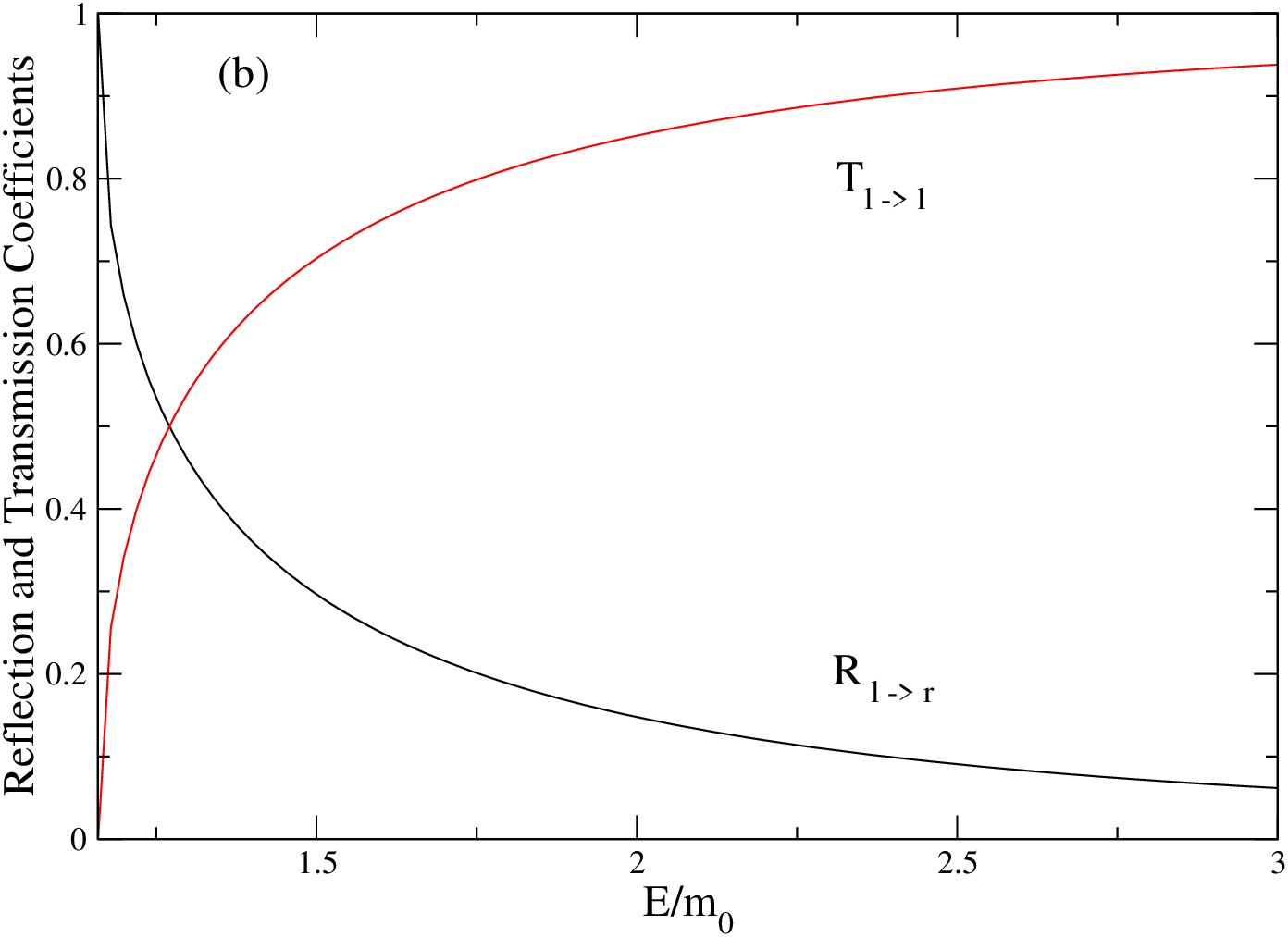}}\par}
\caption{Reflection and transmission coefficients as a
function of energy scaled by the fermion mass for
$B/m_0^2=0.5$, $n=1$, $y_R=4/3$, $y_L=1/3$, $g'=0.344$. (a)
Shows the coefficients for incident, right-handed
helicity modes and (b) for incident, left-handed
helicity modes.}
\end{figure}

The three dimensional treatment of the problem allows for a clearer
physical picture of the fermion motion in the presence of the
external field: Suppose that the original fermion motion
is not parallel to the direction of the field lines and
therefore that its velocity vector contains a component perpendicular
to the field. In this case, due to the Lorentz force,
the particle circles around the field lines maintaining its
velocity along the direction of the field. The motion of the
particle is thus described as an overall displacement along the
field lines superimposed to a circular motion around these
lines. These circles correspond to the different quantum levels, labeled
by $n$, the principal quantum number. We see that the originally
different angles of incidence all result in the same overall direction
of incidence. Also, since the relevant quantities that determine the
strength of the coupling are the magnitudes of the parameters
$\gamma^{R,L}$ and $\gamma$, the generated axial asymmetry is
independent of whether the fermion moves parallel or anti-parallel to
the field lines. 

What is the relation of this axial asymmetry to $CP$ violation? Recall
that in the SM, $CP$ is violated in the quark sector through the mixing
between different weak interaction eigenstates to form states with
definite mass. However, in the present scenario, taking place
within the SM, no such mixing occurs since we are concerned only with
the evolution of a single quark (for instance, the top quark)
species. The relation is thus to be found in the dynamics of the
scattering process itself and becomes clear once we notice that this
can be thought of as describing the mixing of two levels, right- and
left-handed quarks coupled to an external hypermagnetic field. When
the two chirality modes interact only with the external field,
they evolve separately, as described by Eq.~(\ref{finalpsisym}).
It is only the scattering with the bubble wall what allows a finite
transition probability for one mode to become the other. Since the
modes are coupled differently to the external field, these
probabilities are different and give rise to the axial asymmetry. $CP$
is violated in the process because, though $C$ is conserved, $P$ is
violated and thus is $CP$.  

We also emphasize that, under the very general assumptions of $CPT$
invariance and unitarity, the total axial asymmetry (which includes
contributions both from particles and antiparticles) is quantified in
terms of the particle (axial) asymmetry. Let $\rho_i$ represent the
number density for species $i$. The net densities in left-handed
and right-handed axial charges are obtained by taking the differences
$\rho_L-\rho_{\bar{L}}$ and $\rho_R-\rho_{\bar{R}}$, respectively. It
is straightforward to show~\cite{Nelson} that $CPT$ invariance and
unitarity imply that the above net densities are given by
\be
   \rho_L-\rho_{\bar{L}}&=&(f^s-f^b)
   (R_{r\rightarrow l} - R_{l\rightarrow r})\nonumber\\
   \rho_R-\rho_{\bar{R}}&=&(f^s-f^b)
   (R_{l\rightarrow r} - R_{r\rightarrow l})\, ,
   \label{net} 
\ee  
where $f^s$ and $f^b$ are the statistical distributions for
particles or antiparticles (since the chemical potentials are assumed
to be zero or small compared to the temperature, these distributions
are the same for particles or antiparticles) in the symmetric and the
broken symmetry phases, respectively. From Eq.~(\ref{net}), the
asymmetry in the axial charge density is finally given by
\be
   (\rho_L-\rho_{\bar{L}}) - (\rho_R-\rho_{\bar{R}})=
   2(f^s-f^b)(R_{r\rightarrow l} - R_{l\rightarrow r}).
   \label{final}
\ee
This asymmetry, built on either side of the
wall, is dissociated from non-conserving baryon number 
processes and can subsequently be converted to baryon number in the
broken symmetry phase where sphaleron induced transitions are taking
place with a large rate. This mechanism receives the name of 
{\it non-local baryogenesis}~\cite{{Dine},{Nelson},{Cohen},{Joyce}}
and, in the absence of the external field, it can only be realized in
extensions of the SM where a source of $CP$ violation
is introduced {\it ad hoc} into a complex, space-dependent phase of
the Higgs field during the development of the EWPT~\cite{Torrente}.

An interesting possibility to extend the scope of the present work is
to study the scattering of fermions in the presence of topologically
non-trivial configurations of hypermagnetic fields such as the so
called, {\it hypermagnetic knots} which themselves can seed the baryon
asymmetry of the universe in extensions of the SM~\cite{Giovannini2}.  

Since  another consequence of the
existence of an external magnetic field is the lowering of the barrier
between topologically inequivalent vacua~\cite{Comelli}, due to the
sphaleron dipole moment, the use of the mechanism discussed in this
work to possibly generate a baryon asymmetry is not as
straightforward. Nonetheless, if such primordial fields indeed 
existed during the EWPT epoch and the phase transition was
first order, as is the case, for instance, in minimal extensions of the 
SM, the mechanism advocated in this work has to be considered as
acting in the same manner as a source of $CP$ violation that can
have important consequences for the generation of a baryon
number. 

\section*{Acknowledgments}

A.A. Aknowledges useful conversations with G. Piccinelli.
Support for this work has been received in part by DGAPA-UNAM under
PAPIIT grant number IN108001 and by CONACyT-M\'exico
under grant numbers 35792-E and 32279-E.

\end{document}